\documentclass[referee,a4paper,12pt,traditabstract]{jswsc}
\usepackage{graphicx}
\usepackage{subfigure}
\usepackage{epstopdf}
\usepackage[mathlines]{lineno}
\usepackage[authoryear,round]{natbib}
\usepackage[backref]{hyperref}
\usepackage{newtxmath}
\usepackage{mathtools}
\usepackage{url}

\bibpunct[, ]{(}{)}{;}{a}{,}{,}

\hypersetup{colorlinks=true,citecolor=cyan,urlcolor=cyan,linkcolor=blue} 

\begin{document}


\title{Multi-Instrument Observations and Tracking of a Coronal Mass Ejection Front From Low to Middle Corona}

\titlerunning{Multi-Instrument CME Front Tracking}

\authorrunning{Stepanyuk, Kozarev}

\author{O. Stepanyuk
      \inst{1}
      \and
      K. Kozarev\inst{1}
      }

\institute{
          Institute of Astronomy and National Astronomical Observatory\\
          Bulgarian Academy of Sciences\\
          Sofia, Bulgaria\\
          \email{\href{mailto:ostepanyuk@astro.bas.bg}{ostepanyuk@astro.bas.bg}}
          }

\abstract{
The shape and dynamics of coronal mass ejections (CMEs) vary significantly based on the instrument and wavelength used. This has led to significant debate about the proper definitions of CME/shock fronts, pile-up/compression regions, and cores observations in projection in optically thin vs. optically thick emission. Here we present an observational analysis of the evolving shape and kinematics of a large-scale CME that occurred on May 7, 2021 on the eastern limb of the Sun as seen from 1~au. The eruption was observed continuously, consecutively by the Atmospheric Imaging Assembly (AIA) telescope suite on the Solar Dynamics Observatory (SDO), the ground-based COronal Solar Magnetism Observatory (COSMO) K-coronagraph (K-Cor) on Mauna Loa, and the C2 and C3 telescopes of the Large Angle Solar Coronagraph (LASCO) on the Solar and Heliospheric Observatory (SoHO). We apply the updated multi-instument version of the recently developed Wavetrack Python suite for automated detection and tracking of coronal eruptive features to evaluate and compare the evolving shape of the CME front as it propagated from the solar surface out to 20 solar radii. Our tool allows tracking features beyond just the leading edge and is an important step towards semi-automatic manufacturing of training sets for training data-driven image segmentation models for solar imaging. Our findings confirm the expected strong connection between EUV waves and CMEs. Our novel, detailed analysis sheds observational light on the details of EUV wave-shock-CME relations that is lacking for the gap region between the low and middle corona.}

\maketitle

\section{Introduction}
\label{introduction}

Coronal mass ejections (CMEs) are complex eruptive phenomena in the solar corona, most often comprising of a magnetic flux rope core, wrapped in helical magnetic field lines, piled-up coronal plasma, and one or more shock waves (when the CME propagates faster than the local magnetosonic speed) \citep{Vourlidas:2013}. Coronal shocks may also form as blast waves, as a result of the impulsive energy release in flares \citep{Uchida:1974, Vrsnak:1995}. Coronal and interplanetary shocks cause type II radio emission in different wavelength ranges \citep{Cane:1981, ReinerKaiser:1999, WildMcCready:1950, Uchida:1968}.

The first studies of globally propagating waves in the solar atmosphere date back to 1960's when \citet{MoretonRamsey:1960} and \citet{Moreton:1960} observed the phenomena for the first time at the Lockheed Solar Observatory in Burbank and The Sacramento Peak Observatory using time-lapse photography of the solar chromosphere. \citet{Uchida:1968} hypothesized that Moreton-Ramsey waves could be the result of a globally propagating shock wave in the solar corona, but available instruments limited the possibility to investigate further for decades.

After the launch of the Solar and Heliospheric Observatory (SOHO) spacecraft, the EUV Imaging Telescope \citep[EIT]{Delaboudini:1995}
gave opportunity to monitor the full solar disk in 4 extreme UV (EUV) channels with a cadence of $\sim$12~min. Using the Running Difference (RD) technique, \citet{Thompson:1998} observed for the first time a large-scale wave in the low corona with a bright front with a velocity of $\sim$250~km/s, immediately followed by extended dimmings propagating out from the 
flaring site. This was the first of many studies of EIT/EUV waves \citep{Veronig:2010, Kozarev:2011, Long:2014}. RD images are used routinely to study motions in the outer corona. Their main effect is to remove the bright background intensity and enhance the subtle changes of intensity caused by outward (or inward) proper motions of material. \citep{Sheeley:2014} The technique involves taking two consecutive images and subtracting the earlier image from the later one. This results in an image that highlights the changes that occurred between the time when the two original images were taken. In solar imaging, this technique is particularly useful for highlighting transient and fast-moving features. A statistical study on the EIT wave velocity had been performed by \citep{Klassen:2000}. The results show that the velocity varies from 138 to 465~km/s, with an average of 271 km/s. 

Higher cadence studies (2.5 min) with the Extreme Ultraviolet Imager (EUVI) on board the STEREO spacecraft have shown that the EIT wave velocity can be even smaller than the sound speed in the solar corona. Using the Large Angle Spectroscopic Coronagraph \citep[LASCO]{Brueckner:1995} on SOHO, the relationship between EIT (or EUV) waves and CMEs observed in white light was also studied \citep{Tripathi:2007}. In the multiple follow-up studies, these coronal EIT waves appear in EUV as bright fronts propagating across a significant part of the solar disk \citep{Gallagher:2010, Long:2017}.

Originaly seen as coronal counterpart of the chromospheric Moreton-Ramsey waves, attempts were made to explain EUV waves within fast-mode linear wave theory. Discrepancies between the observed and theoretical behavior of the phenomena have led to the development of a series of alternative interpretations for global EUV waves. Discussed mechanisms included stretching of magnetic field lines during a solar eruption \citep{Chen:2002}, continuous small-scale magnetic reconnection \citep{Attrill:2007}, slow-mode magnetohydrodynamic (MHD) waves \citep{Wang:2009}.

Based on the statistical studies, \citet{Biesecker:2002} concluded that EIT waves are more likely to be related to CMEs than to flares. \citet{Chen:2006} showed that regardless of the flare magnitude, EIT waves can be observed only if a CME is present. However, there still exists a dispute on the spatial relation between EIT waves and CMEs. The observational gap between the FOVs of EUV imagers and white-light coronagraphs, in particular, has proven a source of confusion and various interpretations. \citet{Chen:2009} and \citet{Dai:34} found that the EIT wave front is co-spatial with the CME frontal loop, whereas \citet{Patsourakos:2009a} and \citet{Veronig:2010} argued that the EIT wave front is further away from the CME frontal loop.

Limited white-light observations below 3 solar radii and projection effects were an obstacle to understanding early CME dynamics. Several techniques have been developed to address the projection effect issue \citep{Mierla:2008, Thernisien:2009, JoshiSrivastava:2011, HuttonMorgan:2017}, but the implementation of such methods for the inner corona (1.0-1.5~solar radii) has been limited. A three-phase CME kinematic proﬁle takes place due to three separate driving forces -  Lorentz force, the gravitational force, and the viscous drag force, with the latter arising due to interaction with the ambient solar wind \citep{Webb:2012}. Thus, CME kinematics shows three phases - an initial rising phase (weakly accelerated motion \citep{Cheng:2020}, an impulsive phase and a residual propagation phase (constant or decreasing speed - \citep{Gopalswamy:2000}. 

The impulsive phase is characterized by rapid acceleration over a short period of time which gives CMEs high velocities \citep{Cheng:2020, Patel:2021}. Capturing the acceleration phase occurring at the lower coronal heights presents additional challenges, for a numerous reasons: unsuitability of white-light coronagraph observations due to its location \citep{Temmer:2008, Majumdar:2021} and projection effects in measurements in the plane of the sky \citep{Zhang:2006, Balmaceda:2018}.

The launch of the Solar Dynamics Observatory \citep[SDO]{Pesnell:2012}, and the high-resolution, high-cadence EUV telescope Atmospheric Imaging Assembly \citep[AIA]{Lemen:2012}, enabled even more detailed studies of the phenomenon \citep{Kozarev:2015} in the lower corona, as well as extensions to modeling conditions for coronal acceleration of solar energetic particles in compressive fronts \citep{Kozarev:2016, Kozarev:2017, Kozarev:2019}. Higher up, the ground-based COronal Solar Magnetism Observatory's \citep[COSMO]{Tomczyk:2022} K-coronagraph (K-Cor) on Mauna Loa was specifically designed to study the formation and dynamics of coronal mass ejections and the evolution of the density structure of the low corona. It has been used for regular observations of the corona and its activity. Its FOV spans well the EUV-white light observational gap, its main shortcoming being the lower signal to noise ratio due to the atmosphere. Nevertheless, this instrument holds much potential for characterising eruptive features, and has even been proposed as a near-real time CME monitoring tool, to be used for forecasting SEPs \citep{StCyr:2017}.

In recent years Machine Learning (ML) methods have become more frequently applied in solar physics. For instance \citet{Szenicer:2019} used a Convolutional Neural Network (CNN) to produce EUV irradiance maps from AIA images. \citet{Li:2013} proposed a multi-layer model to predict the solar flare based on sequential sunspot data; and \citep{Kim:2019} applied generative adversarial networks (GAN) to generate the magnetic flux distribution of the Sun from SDO/AIA image. Nevertheless, usage of data-driven approaches for tracking of CME-related phenomena is currently limited due to insufficiency of training sets. 

Recently, we proposed a method for smart characterization and tracking of solar eruptive features \citep{Stepanyuk:2022}, and have shown its performance on a small set of CME-related phenomena observed with the AIA telescope. One of the purposes of that work was to create a tool that would make preparing CNN training sets a more easy task. Here we extend this concept and present an updated version of the software with multi-instrument observational analysis of the evolving shape and kinematics of a large-scale CME that occurred on May 7, 2021 on the eastern limb of the Sun as seen from 1 AU. The eruption was observed continuously, consecutively by AIA, K-Cor, and the LASCO C2 and C3 telescopes.
Most of the existing CME analysis tools focus just on leading edges and fronts ignoring the dynamics of the interior of the CME. Our method addresses this problem, separately allowing to highlight both the internal development and the leading edges of the features.

The paper is structured as follows: We describe the method briefly in Section \ref{method}. In Section \ref{observations} we describe the various observations. In Section \ref{analysis} we present an analysis and discussion of the evolution of the eruption based on the multi-wavelength, multi-instrument observations. We provide conclusions in Section \ref{conclusions}.

\section{Method}
\label{method}

\subsection{Feature Recognition}

Data from the 193~\AA~channel of the AIA instrument was used to study the early propagation of the eruption and of the associated EUV front in the lower corona. It was downloaded and processed to level 1.5 using the standard AIA processing pipeline in the Python aiapy library \citep{Barnes:2020}. The K-Cor coronagraph records the polarization brightness (pB) formed by Thomson scattering of photospheric light by coronal free electrons. Its field of view spans from 1.05 to 3 solar radii. The nominal recorded images have a size of 1024$\times$1024 pixels, with 5.6~arcsec/pixel resolution and 15-second cadence. The K-Cor observations used in this work were 2-minute averaged normalising-radial-graded-filter \citep[NRGF]{Morgan:2006} images, which have been processed with the K-Cor standard pipeline, and downloaded from the Mauna Loa Solar Observatory's web page \footnote{https://doi.org/10.5065/d69g5jv8}. LASCO C2 and C3 level-1 data, calibrated and processed for instrumental response, were obtained using the Sunpy Python library \citep{Sunpy:2020}.

We applied an updated, multi-instrument version of the recently developed Wavetrack code for automated detection and tracking of coronal eruptive features \citep{Stepanyuk:2022} to evaluate and compare the evolving shape of the CME front as it propagated from the solar surface out to 20 solar radii. Initially, the framework performs wavelet decomposition of observational data and applies one or a few filtering techniques to each of the wavelet decomposition levels, with further recomposition and segmentation of the resulting image. The framework follows a modular approach and is built as a set of classes where each class represents one or a few image processing techniques, so that these classes act as building blocks allowing various decomposition and processing configurations and setups, depending on the input image characteristics. The scheme on Figure 2 in \citet{Stepanyuk:2022} gives a general overview of the image processing stages with the Wavetrack software. 


The specific Wavetrack processing for this event goes as follows. Base difference images were obtained by subtracting a base image from each timestep of the image data. Using them allows to enhance the change in intensity, caused by the eruptive front, omit static details, and reduce noise. Base images are created as an average of several (3-5) consecutive timesteps 2-5 minutes prior to the beginning of the event in each instrument. These were determined by visual inspection. For the AIA instrument only, absolute values of the threshold interval (-50,150) are selected for the purpose of narrowing of the image dynamic range. For AIA and LASCO C2/C3 instruments, base difference images were reduced to 8 bit depth, 16 bits were used for K-Cor instrument. At the next step, base difference images were decomposed with the \`a trous wavelet technique into a series of scales, with third and fourth wavelet scales chosen for image recomposition for AIA instrument. To each of the wavelet coefficients a relative thresholding is applied once more depending on the statistical distribution of the pixel intensities for each of the decomposition levels. Finally, after segmentation of the object masks at each time step they are multiplied by the original data to reveal the intensity of different parts of the object \citep{Stepanyuk:2022} and provide data for velocity field estimation. More detailed parameters of processing, decomposition and filtering techniques are given in the examples section of the Wavetrack package\footnote{\url{https://gitlab.com/iahelio/mosaiics/wavetrack}} for the LASCO, K-Cor and AIA instruments.

\subsection{Solar Eruptive Feature Velocity Estimation}
Velocity maps provide a spatially-resolved depiction of velocities across a feature or region in the plane of sky, allowing for a clear visualization of how velocities vary from one part of the feature to another. There are a number of velocity field estimation approaches, each with its own precision and sensitivity, and thus expected to perform differently depending on the instrument data characteristics and quality. For this reason, in the new version of the Wavetrack software we implemented Fourier Local Correlation Tracking (FLCT) \citep{welsch_flct:2004},  Horn-Schunk \citep{horn:1981} and Lukas-Kanade \citep{lucas:1981} method.

In solar physics, the FLCT method has emerged as a popular technique for determining plane-of-sky flows on the Sun's surface, particularly in active regions \citep{loptien_b_flct:2016, mckenzie_flct:2013, su_flct:2013}. By tracking the movement of magnetic features in successive solar images, FLCT provides a detailed map of solar surface velocities. This has proven crucial for studying phenomena like the emergence and evolution of sunspots, magnetic reconnection processes, and the dynamics of solar flares. It has also been used for large-scale feature tracking \citep{Telloni:2022} The FLCT method utilizes local correlation tracking combined with a Fourier technique to enhance accuracy and reduce noise. 

Optical flow is a fundamental concept in computer vision and image processing. It refers to the apparent motion of brightness patterns in the image. A few optical flow methods were added to the Wavetrack package, aiming to provide choice while analyzing data coming from different instruments with varying cadence.

The Lucas-Kanade method \citep{lucas:1981} is a widely-used differential method for optical flow estimation developed by Bruce D. Lucas and Takeo Kanade. It assumes the flow is essentially constant in a local neighborhood of the pixel under consideration, and uses the local image gradients and temporal gradients to solve for the optical flow parameters. The method works well for small motions and is more accurate in estimating finer details.

The Horn-Schunck Method \citep{horn:1981} is global in nature, meaning it uses equations based on brightness constancy over the entire image. It introduces a smoothness constraint that penalizes non-smooth flow fields, aiming to capture more global motion. The regularization term (smoothness constraint) makes it robust to noise but can sometimes over-smooth the result, missing finer motion details.
%

Still, there is space for ambiguity when it comes to the discussion on how the velocity of a solar eruptive feature is supposed to be defined. Features like EUV waves, active regions, and to a lesser extent filaments generally change shape as they evolve, while different parts of an object can move with different velocities themselves and pixel intensity values of corresponding regions can vary between cadence steps. That makes it logical to also operate with some general integral quantity, such as center of an object mass (for more details, see Section \ref{flct_results}).

\section{Observations}
\label{observations}

The event on May 07, 2021 began at 18:43~UT with an M3.9-class flare located at [N17,E78] on the solar disk, as reported by the Heliospheric Event Knowledgebase \citep[HEK]{Hurlburt:2012}. The flare originated in AR 12822, which was turning around the eastern limb as observed from the Earth, and was also the source of the EUV wave. AIA first observed it at 18:55~UT. The EUV front developed into a dome-like shape, similar to many previously reported large-scale EUV waves. Being relatively close to the solar equator, its nose quickly reached the edge of the AIA field of view. The off-limb southern flank of the front was more pronounced and brighter, as was the on-disk signature. The left panel halves in Figure \ref{fig_wavetrack_aia} show the EUV wave evolution in base difference images for six epochs, spanning $\sim$8 minutes.

The front was first seen in white light in the K-Cor coronagraph at 18:56~UT, propagating in the east-northeast direction, into a streamer region. It exhibited the same dome-like shape as the EUV feature, but with an uninterrupted front, increasing in thickness with time. Behind the front a dimming region developed, and a small but bright core was visible, possibly the driver. The left panel halves in Figure \ref{fig_wavetrack_kcor} show the eruptive front in base difference images for six epochs between 19:02~UT and 19:23~UT, when it reached the edge of the K-Cor FOV.

In LASCO C2 the eruption was first observed at 19:24~UT off the eastern limb. The CME reached the edge of the C2 FOV at 20:28~UT. Figure \ref{fig_wavetrack_lascoc2}, left half-panels, shows the integral images at four times, from 19:24~UT to 20:28~UT. The eruption retained its dome-like shape for the first two observations, and became distorted already by 19:48~UT, possibly from drag exerted on it by the faster solar wind as it exited the streamer region off the northeastern limb (clearly seen in the integral image at 19:24~UT).

In LASCO C3 the CME front first appeared at 19:54~UT. By 20:34~UT a single leading front is no longer visible, similar to the 20:28~UT C2 observation. Over time, the CME divides into a southern and northern feature. This is visible in Figure \ref{fig_wavetrack_lascoc3}, left half-panels. We follow it until 22:30~UT; by then the sky-projected front of the CME is at nearly half the distance to the edge of the FOV, or approximately 16 solar radii.


\section{Analysis}
\label{analysis}

\subsection{Wavetrack results}
\label{wavetrack_results}

\begin{figure}
\centerline{\includegraphics[width=0.7\columnwidth]{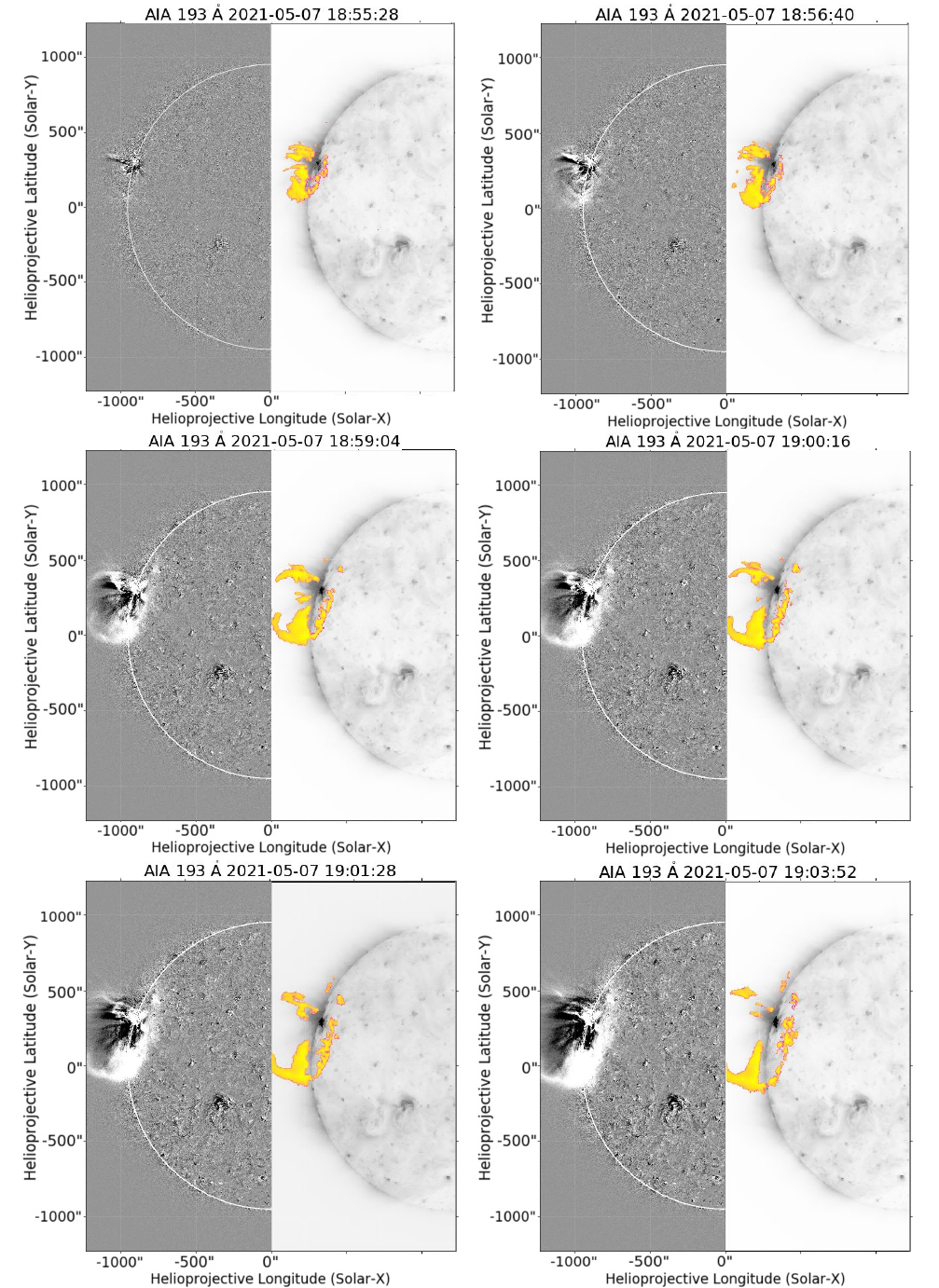}}
    \caption{Six split panels showing the development of the EUV wave in base difference AIA images (left sides) and the Wavetrack-detected features (right sides).}
    \label{fig_wavetrack_aia}
\end{figure}

\begin{figure}
\centerline{\includegraphics[width=0.7\columnwidth]{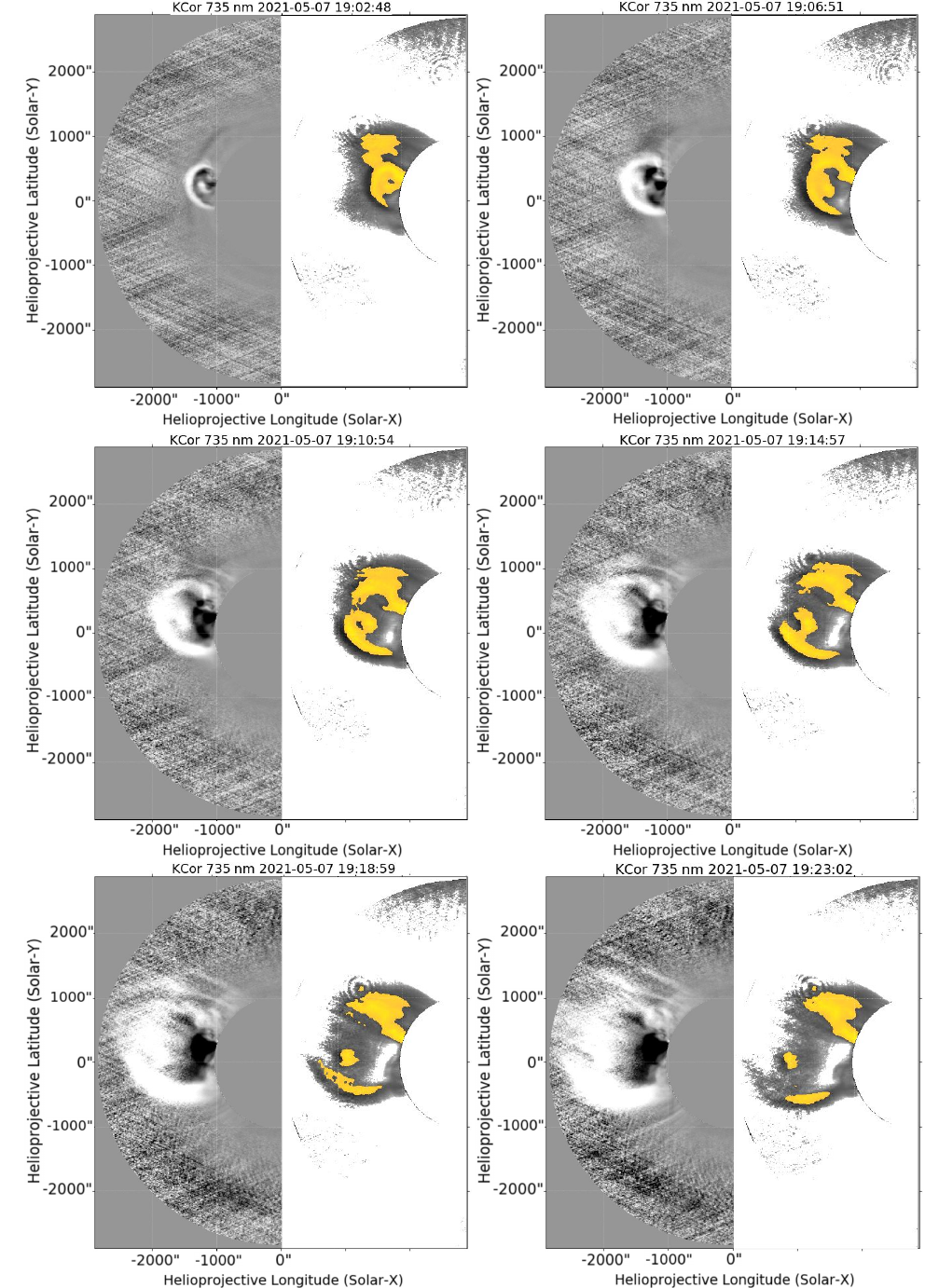}}
    \caption{Same as Fig. \ref{fig_wavetrack_aia}, but for the COSMO K-Cor coronagraph observations.}
    \label{fig_wavetrack_kcor}
\end{figure}

\begin{figure}
\centerline{\includegraphics[width=0.7\columnwidth]{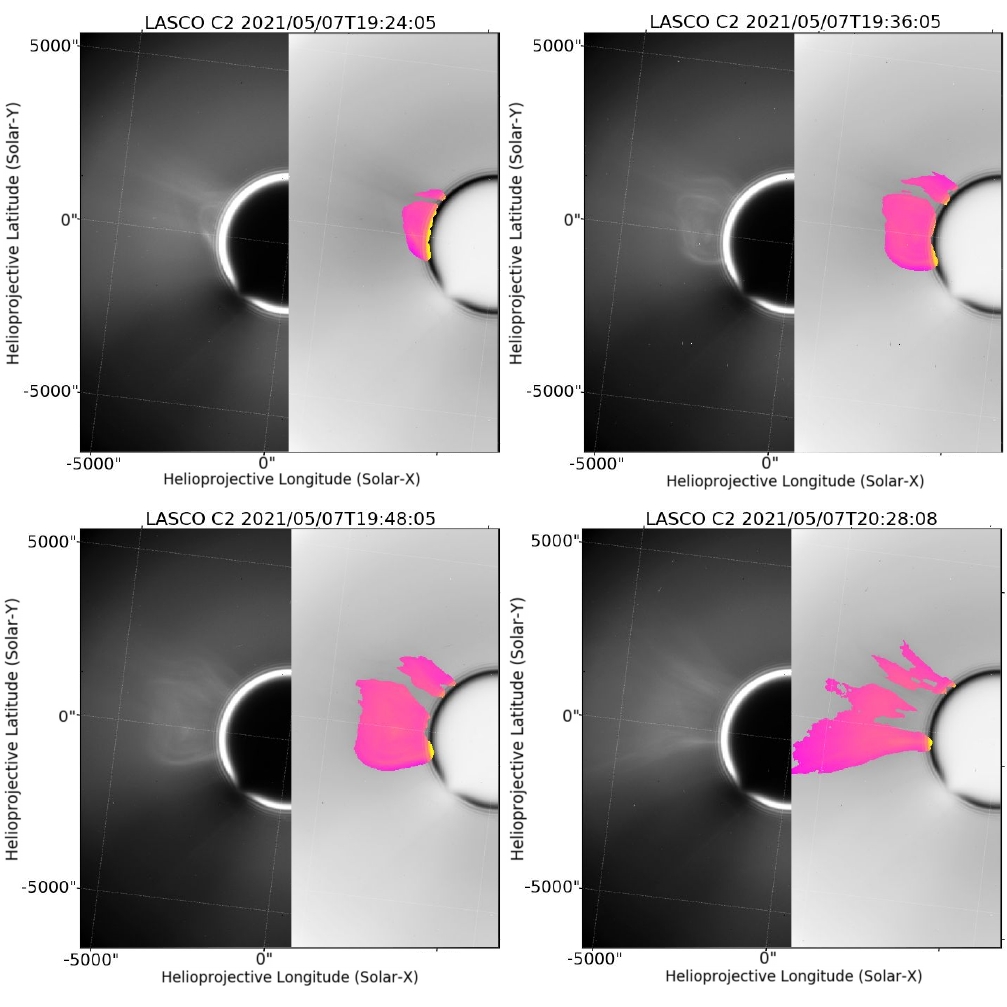}}
    \caption{Same as Fig. \ref{fig_wavetrack_aia}, but for the LASCO C2 coronagraph observations. The left sides show integral LASCO C2 images.}
    \label{fig_wavetrack_lascoc2}
\end{figure}

\begin{figure}
 \centerline{\includegraphics[width=0.7\columnwidth]{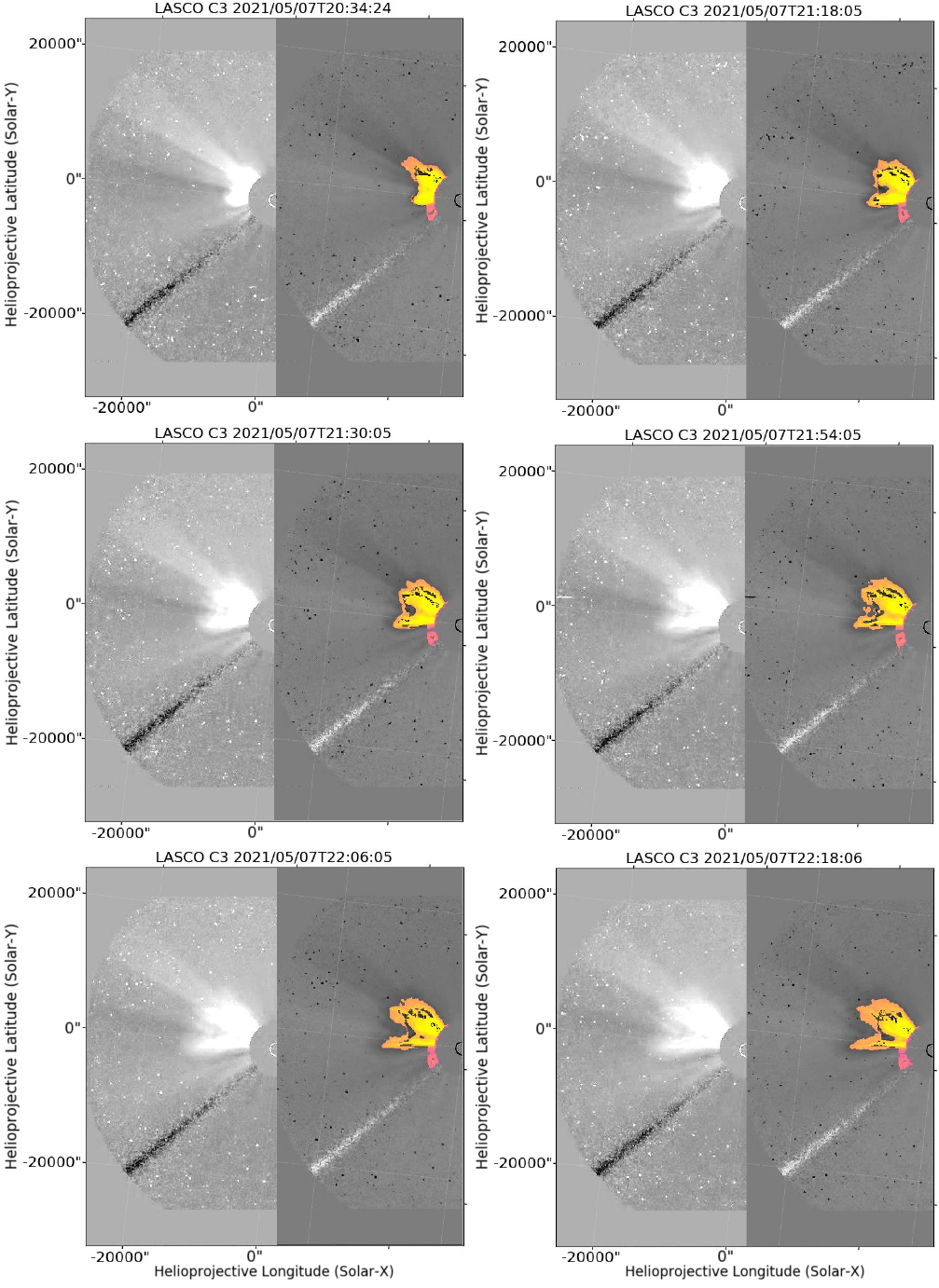}}
    \caption{Same as Fig. \ref{fig_wavetrack_aia}, but for the LASCO C3 coronagraph observations.}
    \label{fig_wavetrack_lascoc3}
\end{figure}

We performed separate Wavetrack processing for the AIA, K-Cor, LASCO C2, and LASCO C3 images. The algorithm captures well the large-scale shapes of the observed features. The right sides of the panels in Figures \ref{fig_wavetrack_aia}-\ref{fig_wavetrack_lascoc3} showcase the results: fully segmented features colored in yellow and magenta for selected observation epochs, overlaid on the original data in greyscale (with regular or inverted color maps for better contrast). The figures allow direct comparison with the corresponding base difference and integral images for each instrument. The colored feature pixels are the result of pixel-wise multiplying of the final binary masks of Wavetrack with the original integral data of the observations. Thus, they contain the intensity variation of the features.

Figure \ref{fig_wavetrack_aia} (right half-panels) shows that the feature starts as a typical EUV wave in the AIA low corona FOV: a global dome developing gradually, propagating in all directions from the flare/AR source. In this case it propagated radially outward and to the south, starting from the AR. The front is thinnest in the radial propagation direction, which may be driven. It is better defined off-limb to the south of the nose, with a significant and long-lasting brightening behind it. This is the predominant non-radial propagation direction. On the solar disk, the EUV wave signature is easily seen propagating away from the flaring AR, and matches the off-limb feature position in the south.

Unlike the EUV feature, in white light the front retains its early dome-like shape, as it develops in the FOV of K-Cor (Fig. \ref{fig_wavetrack_kcor}, right half-panels). The northern flank is mostly stationary, rooted in the flaring AR. Eventually, the southern flank overtakes and separates from the rest of the feature, as the front opens at the nose. The Wavetrack processing has also captured the driver in the radial direction, shown as a separate small blob in the center of the CME.

Further away from the Sun, the LASCO C2 white-light feature is well detected and tracked by Wavetrack (Fig. \ref{fig_wavetrack_lascoc2}, right half-panels). The dome-like front retains its shape near the nose only until 19:36~UT. After that only the flanks remain easily discernible. The eruption reaches the edge of the C2 FOV around 20:28~UT.

Finally, the eruptive feature is also tracked by Wavetrack in LASCO C3 observations (Fig. \ref{fig_wavetrack_lascoc3}, right half-panels). The CME is quite different from the initial AIA-observed feature, likely affected by the rapidly changing coronal plasma density and magnetic field, and deformed by the interaction with large-scale coronal structures and the nascent solar wind. The observed feature does not change significantly in shape over the 2.5~hours of tracking in the C3 FOV.

\subsection{Wavetrack Features Comparison}
\label{wavetrack_feature_comparison}

\begin{figure}
\centerline{\includegraphics[width=0.7\columnwidth]{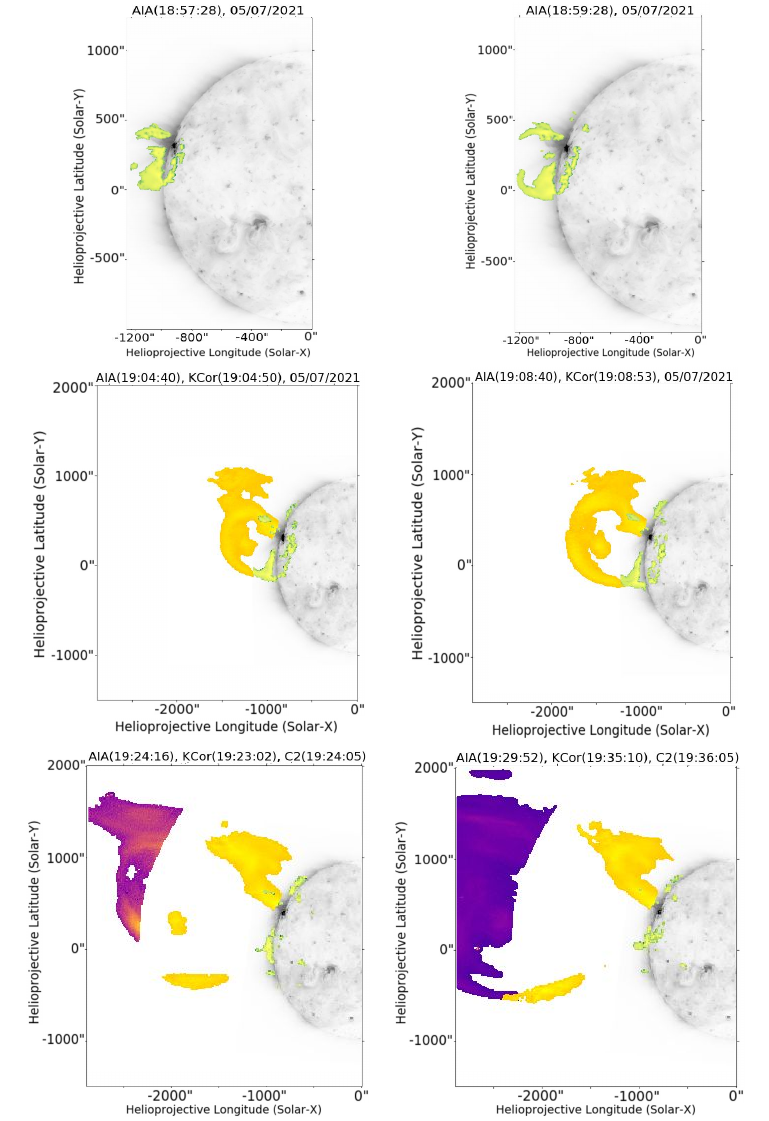}}
    \caption{A comparison of the feature morphologies from the different instruments. Green is AIA, yellow - K-Cor, Purple - LASCO C2.}
    \label{fig_wavetrack_multi_instrument}
\end{figure}

A significant advantage of the Wavetrack tracking method is its ability to extract the features at different times or between different observing sources, and to compare them directly. We show in Figure \ref{fig_wavetrack_multi_instrument} the evolution in multi-instrument observations of the May 07, 2021 eruptive front in six panels, which include Wavetrack features from AIA, K-Cor, and LASCO C2, between 18:57~UT and 19:24~UT. The front is not visible in the LASCO C3 observations. The first two epochs (top two panels) only show the AIA features (in green) early on, overlaid on the corresponding integral 193-channel images with inverted greyscale color map at the same time. The middle panels show K-Cor features (in yellow) added to the AIA features, again overlaid on integral AIA images. The bottom two panels also show the C2 features, while AIA on-disk signatures are included to demonstrate the connection to the off-limb front.

By overlaying the features obtained in our method, we can follow closely the CME morphology over time, taking advantage of the availability of K-Cor data that bridges the low corona observational gap. This also allows us to relate directly the EUV wave and white-light CME. In the standard observational paradigm, the AIA feature should be a compressive MHD wave, while the white-light features from K-Cor and LASCO C2 are expected to show the CME flux rope and core. We find here that while the AIA feature is a classic EUV wave, there is an excellent match between the observed eruptive front in EUV and white-light observations, as seen in Fig. \ref{fig_wavetrack_multi_instrument}. Either the EUV feature is related to expanding magnetic loops, or the white-light features are related to a compressive front. We claim the latter occurred in this event, supported by two observational findings. 1) The white-light front in K-Cor is well defined, and it becomes thicker in time. 2) The flanks of the K-Cor feature, which match well with the LASCO C2 feature flanks, also match very well with the expanding on-disk AIA feature, i.e. the on-disk EUV wave. We thus propose the observed feature in K-Cor is actually a well defined pile-up compression region behind a compressive or shock wave in the low corona, which has not yet detached. This rare observation thus lends further support to the wave nature of coronal bright EUV fronts.

\subsection{Velocity Calculation Results}
\label{flct_results}
In addition to the morphology of the May 07, 2021 eruption, we studied its kinematics, projected onto the plane-of-sky.

To study the kinematics of the coronal front features in detail, initially we employed the Fourier Local Correlation Tracking (FLCT) method \citep{welsch_flct:2004, Fisher:2008}. When applied to the object masks output of Wavetrack, it produces detailed maps of the instantaneous velocity of the object of interest. In this application, initially we use a freely available Python implementation of the method as a Sunpy-affiliated package\footnote{\url{https://github.com/sunpy/pyflct/}}.
The software produces instantaneous plane-of-sky velocities to each data pixel in consecutive images. Specifically, it provides X- and Y-velocity components, allowing to show the magnitude and direction of the evolving features in detail - including expansion. In this study, we applied it to the Wavetrack features obtained for each instrument's observations. More details and various aspects of using FLCT with Wavetrack are described in \citet{Stepanyuk:2022}. We discuss below the individual results and how they compare among the instruments.

\begin{figure}
\centerline{\includegraphics[width=0.7\columnwidth]{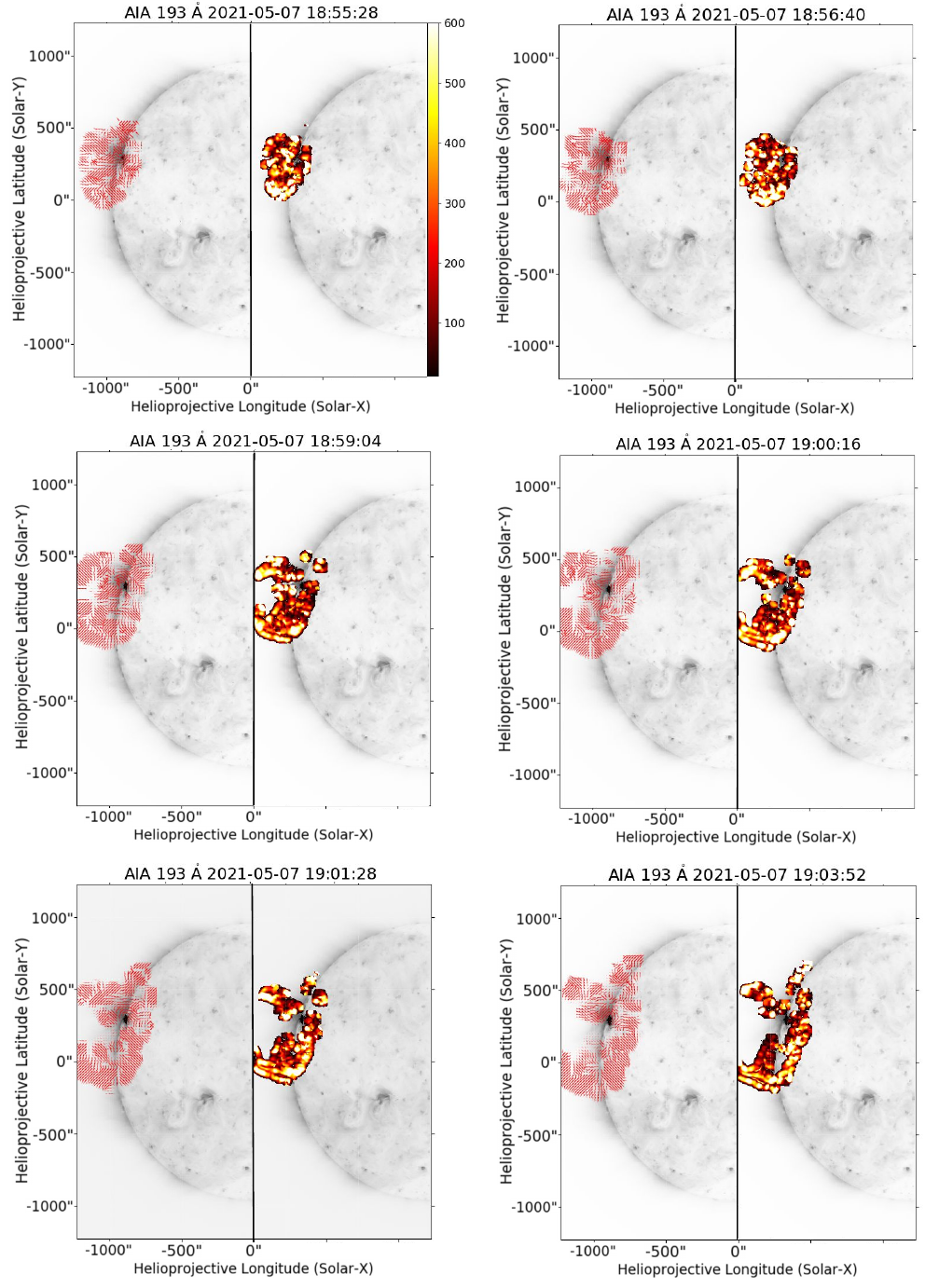}}
    \caption{For the same times as in Fig. \ref{fig_wavetrack_aia}, the corresponding FLCT speeds (right half-panels) and the velocity vectors (left half-panels).}
    \label{fig_flct_aia}
\end{figure}

Figure \ref{fig_flct_aia} shows the FLCT velocities for the AIA feature, for the same six times as Fig. \ref{fig_wavetrack_aia}. Each panel is split in two - on the left, red arrows denote the direction of evolution of the EUV wave. The direction is shown only for a subset of the pixels, in order not to crowd the figure. On the right, the plane-of-sky speeds for each feature pixel are shown in color coding. As expected, the most dynamic parts of the EUV wave have the highest speeds - the northern and southern flanks that expand above the eastern solar limb reach instantaneous speeds of over 500~km/s. The on-disk part of the transient also reaches similar high speeds as it propagates away from the flaring AR. The central and northern on-disk parts of the feature show lower speeds of up to 400~km/s. The directions of motion within the feature (left half-panels) confirm the extensive southward expansion.

\begin{figure}
\centerline{\includegraphics[width=0.7\columnwidth]{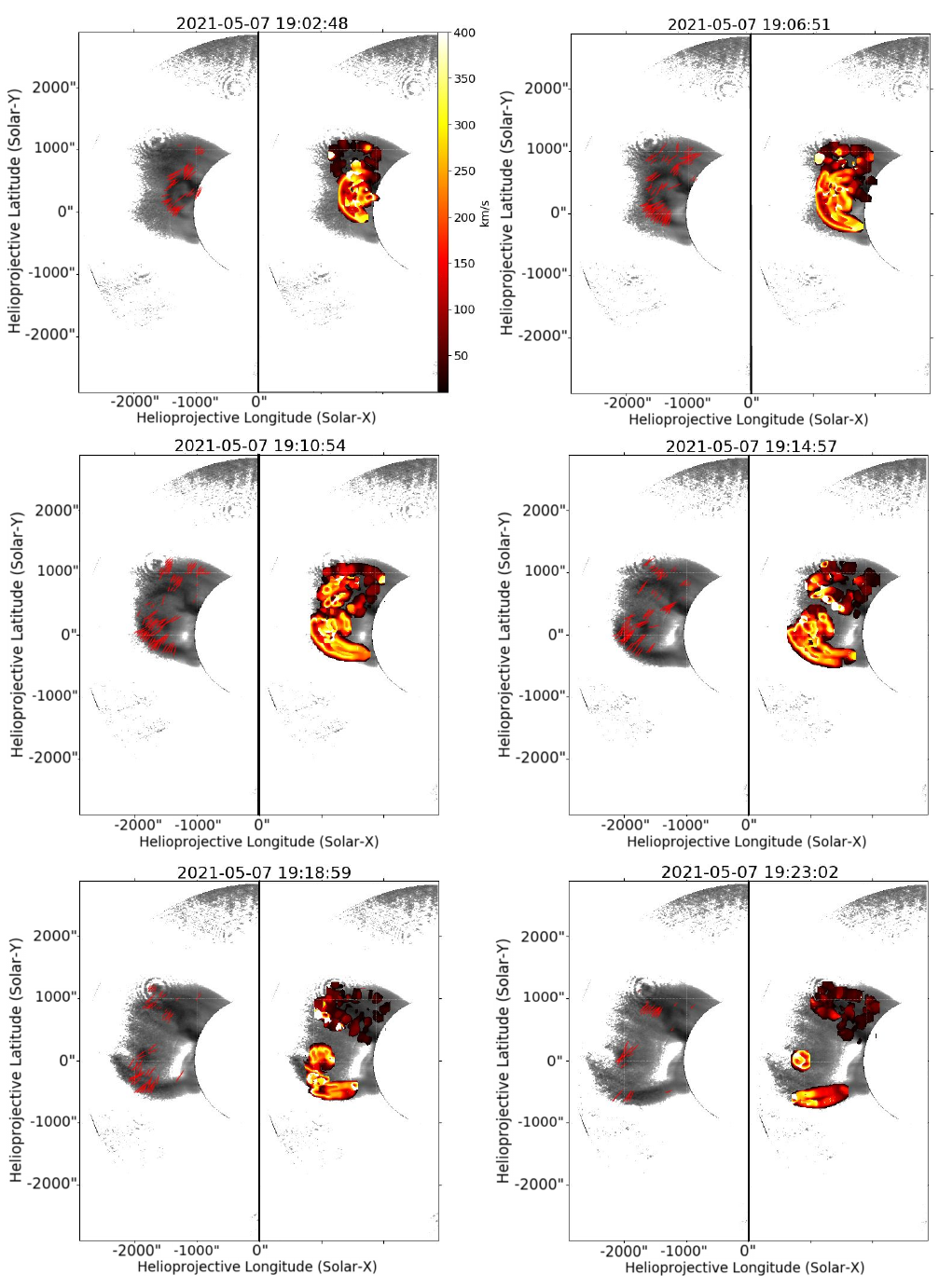}}
    \caption{Same as Fig. \ref{fig_flct_aia}, but for the COSMO K-Cor coronagraph observations.}
    \label{fig_flct_kcor}
\end{figure}

In Figure \ref{fig_flct_kcor}, we show the FLCT speeds and directions of motion for the K-Cor white-light eruptive feature. As seen in the figure, initially the entire front is expanding with speeds up to 400~km/s, but eventually the plane of sky speeds of the northern part of the feature are reduced to lower than 200~km/s, and only the southern flank of the feature continues moving with higher speeds. The small driver feature behind it is also moving fast throughout the period of observation. This behavior is consistent with the plane-of-sky velocities in the AIA field of view, though somewhat lower. 

\begin{figure}
\centerline{\includegraphics[width=0.7\columnwidth]{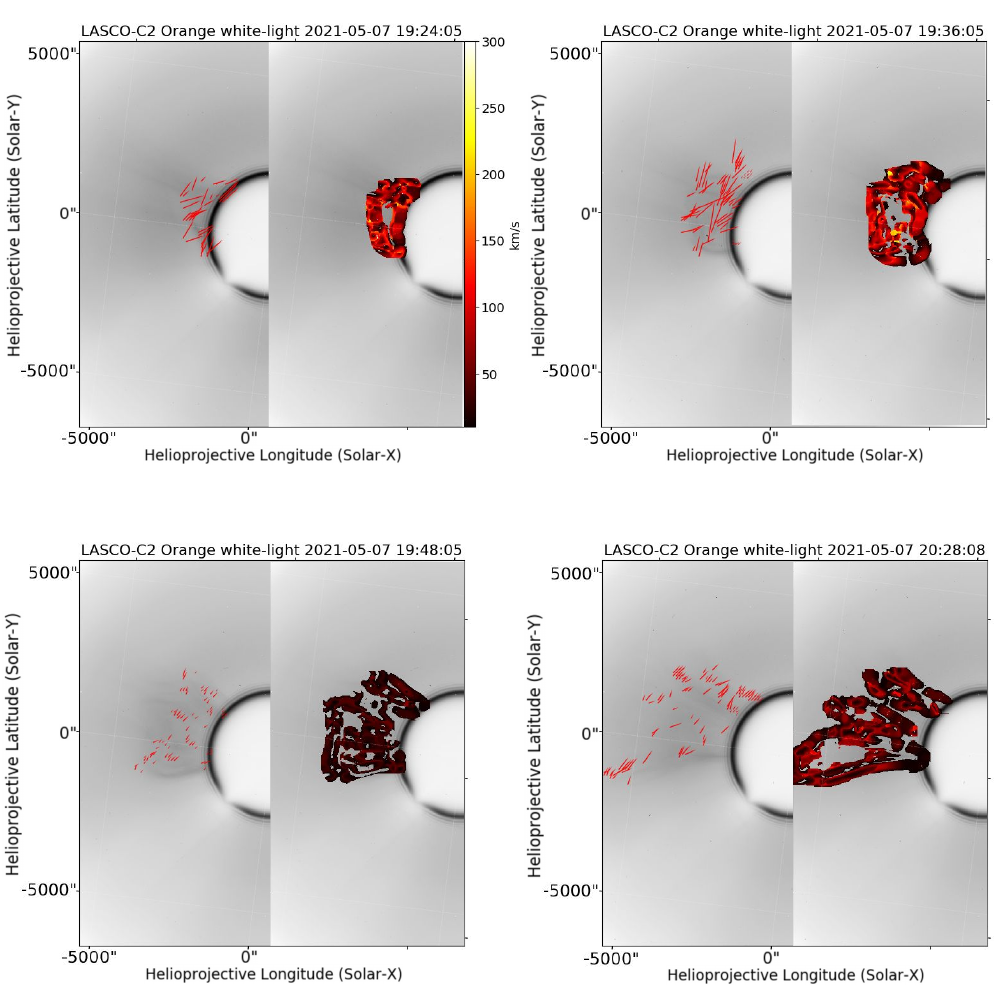}}
    \caption{Same as Fig. \ref{fig_flct_aia}, but for the LASCO C2 coronagraph observations.}
    \label{fig_flct_lascoc2}
\end{figure}

Figure \ref{fig_flct_lascoc2} shows the LASCO C2 speeds as determined by the FLCT method.  Initially, the speeds were between 150-200~km/s along most of the fronts, but later in the event declined to 100~km/s or less, in this estimation. The tangential expansion was also only evident early in the event, judging by the velocity vectors (left half-panels). We note there is a significant gap between the third and fourth LASCO C2 observation due to a switch in the observing regime. Low velocity values obtained with FLCT method can be explained by the effect of sampling cadence. When presented with relatively low sampling of the data, it is generally found that the simplest differentiation techniques are not applicable \citep{byrne:2013}. In this particular case, additional complexity came from the input data with two consecutive timesteps missing (variable cadence). Control calculations with Wavetrack FLCT showed similar results.

To improve the estimation of the velocity field for the LASCO C2 instrument, we applied two other methods for plane-of-sky velocity estimation: the Lukas-Kanade (LK) and Horn-Schunk (HS) algorithms. 
The LK method provides sufficient detail, however it also gives similarly low speeds as FLCT, and may suffer from the same issues for the LASCO C2 data.

%

\begin{figure}
\centerline{\includegraphics[width=0.7\columnwidth]{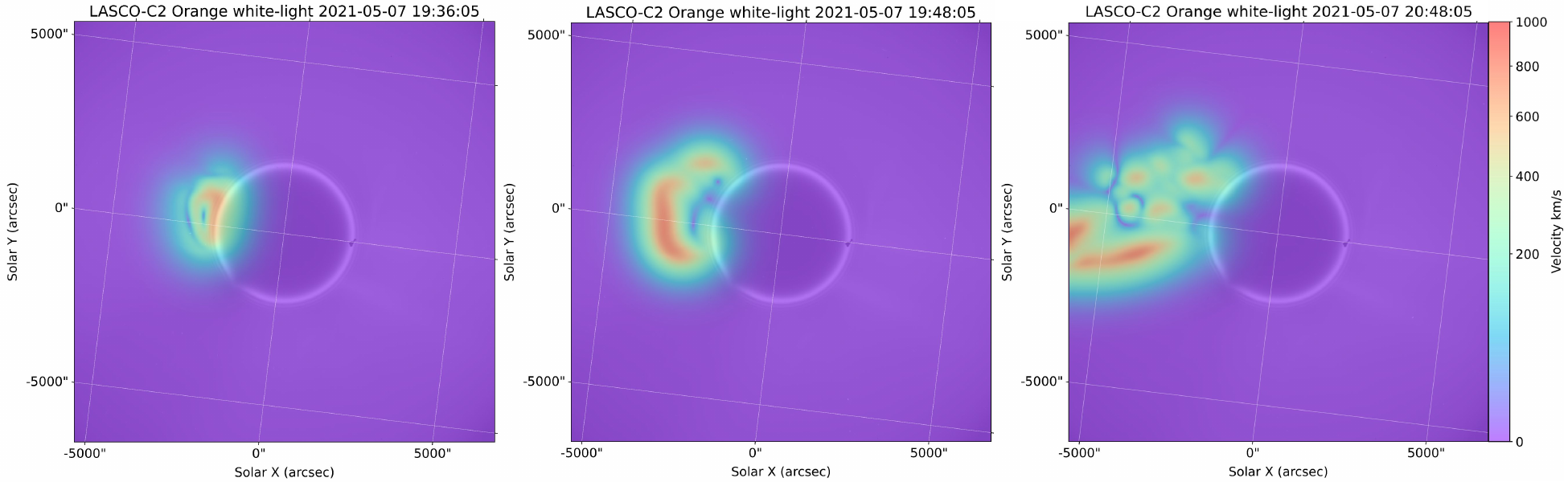}}
    \caption{Velocity field as calculated by Horn-Schunk algorithm for the LASCO C2 coronagraph observations.}
    \label{fig_hs_lascoc2}
\end{figure}

The HS method application is shown in Figure \ref{fig_hs_lascoc2}. It provides less detail and a more large-scale picture, which in the case of the global C2 observations is acceptable. The application shows that pixels of the most dynamic parts of the features show speeds of 470-850~km/s. Initially, the expansion close to the Sun is fastest (left panel). Then the leading front near the nose has the highest speeds (shown in the middle panel), but later on (third panel) the southeastern flank is fastest, consistent with significant overexpansion in that direction.

The different performance of the velocity methods required additional investigation. We found that the HS results are consistent with records in the  CACTus\footnote{\url{https://www.sidc.be/cactus/catalog.php}} and CDAW\footnote{\url{https://cdaw.gsfc.nasa.gov/CME_list/}} CME catalogs, which for this CME provide speed estimates of 754~km/s and 625~km/s, respectively, based on the front motion alone. Additionally, we estimated the object velocity by calculating the speed of the center of mass of the C2 Wavetrack features for at least two consecutive time steps. The center of mass is calculated based on pixel coordinates with pixel intensities taken as weights. The feature center of the mass reaches speeds of 900-1100 km/s in the eastern direction for LASCO C2 Instrument data. Thus, we are more confident in the application of the HS algorithm to the C2 features in this event than in the results from the FLCT and LK methods.

\begin{figure}
    \centerline{\includegraphics[width=0.7\columnwidth]{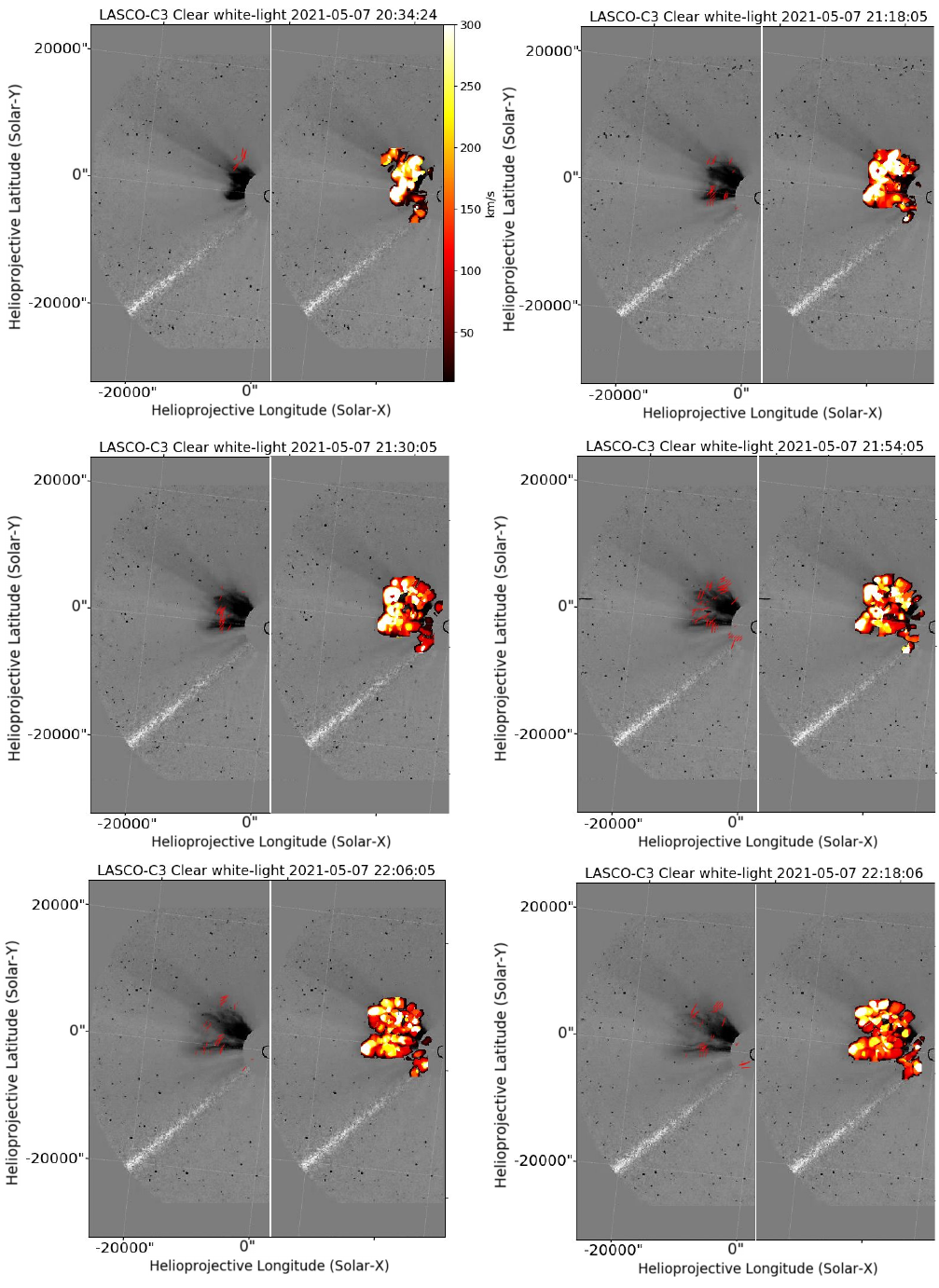}}
    \caption{Same as Fig. \ref{fig_flct_aia}, but for the LASCO C3 coronagraph observations.}
    \label{fig_flct_lascoc3}
\end{figure}

Finally, Figure \ref{fig_flct_lascoc3} shows the LASCO C3 speeds as determined by the FLCT method. We note that the CME speeds have increased farther from the Sun, although the front shape is not visible anymore. Speeds higher than 250~km/s are common in all panels, especially towards the nose and northern/southern expanding `tongues'. This may be due to the solar wind picking up speed and accelerating the CME, which is apparent in the rapid change of its morphology. There is hardly any lateral expansion of the front of the feature.

\begin{figure}
\centerline{\includegraphics[width=0.8\columnwidth]
{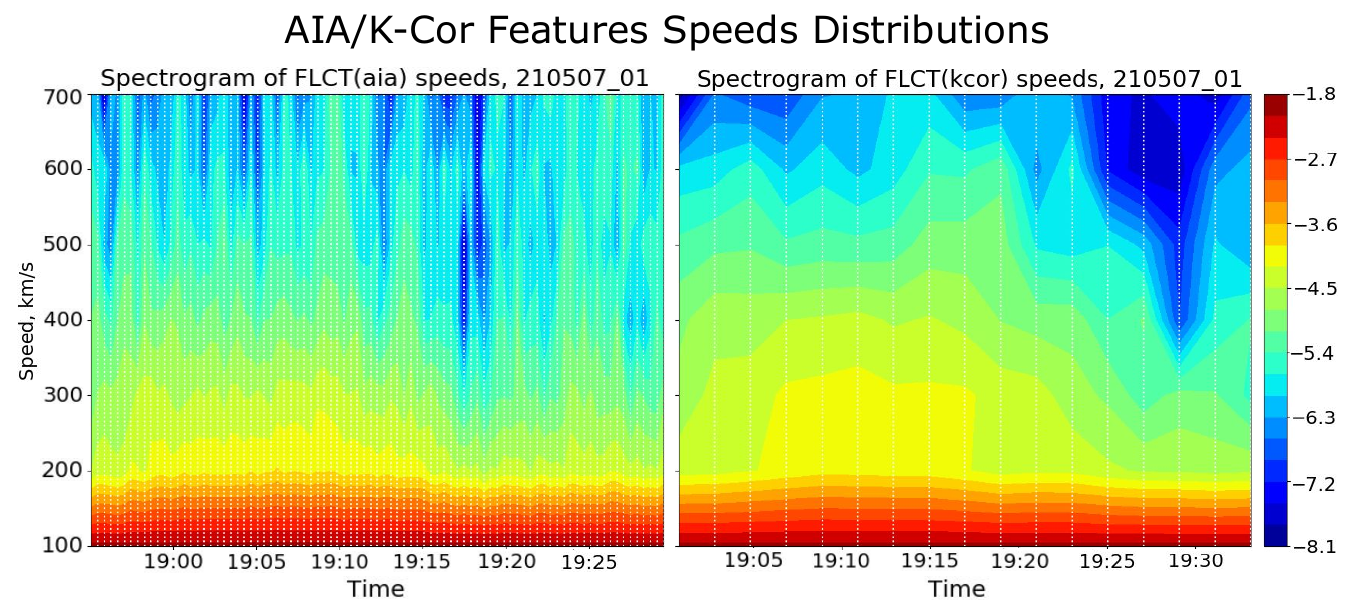}}
    \caption{Spectrogram plots showing the plane-of-sky speed distributions for the Wavetrack features estimated with the FLCT method in the four instruments. Time is on the X-axis, speed on the Y-axis, and the color coding represents the logarithm of the histogram density values.}
    \label{fig_flct_speeds_comparison}
\end{figure}


We also compared the FLCT-estimated speeds over time in AIA and K-Cor instruments. Figure \ref{fig_flct_speeds_comparison} shows distributions over all image pixels of the FLCT-estimated speeds over the eruption duration. Separate panels are shown for each of the two instruments. We calculated normalized pixel distributions for the feature speed values, and show the logarithm of the histogram. In order to compare the histograms calculated for different total number of pixels in the AIA and K-Cor images, we computed and show histogram densities rather than total pixel numbers. The color table represents the value of the logarithm of histogram density. Vertical dotted lines denote the observation times.

We find overall agreement in the temporal behaviour of the speed distributions between AIA and K-Cor. In particular, in both histograms a gradual increase and then decrease in the histogram density for higher-speed pixels ($>$200~km/s) is observed ($-5<\log(density)<-4$ or green-yellow colors in Fig. \ref{fig_flct_speeds_comparison}). LASCO C2 speeds estimated with FLCT are significantly lower,  due to issues of the algorithm when presented with a small number of images and a variable cadence, and therefore LASCO data is not included in Fig. \ref{fig_flct_speeds_comparison}.


\section{Conclusions}
\label{conclusions}

In this work we present a rare continuous observation the early evolution of a coronal eruptive feature that occurred on May 07, 2021, focusing on its appearance in different wavelengths and instruments. We performed automated detection, tracking, and analysis of an EUV wave and related white light CME front using the SDO/AIA, COSMO K-Cor, LASCO C2, and LASCO C3 instruments. The updated Wavetrack method tracks in detail the major dynamic features, allowing to extract the front morphology over time. We find significant overlap of the front geometry first between the AIA and K-Cor dynamic features, and then between the K-Cor and LASCO C2 features, as tracked by Wavetrack. The angular extent of the features over time is similar between the four instruments throughout the event. Fourier Local Correlation Tracking analysis provided 2D-resolved estimates of the coronal eruption velocities over time, and show an overall consistency between the plane-of-sky speeds of the features in the four instruments - especially between the AIA and K-Cor observations. The estimated feature speeds vary as a function of position along the eruptive front. In regions of strong expansion and propagation early on, speeds reached over 500~km/s. Later in the middle corona the CME speeds exceed 700 km/s, consistent with CME catalogs.

Our findings confirm the expected strong connection between EUV waves and CMEs. Our novel, detailed analysis sheds observational light on the details of EUV wave-shock-CME relations that is lacking for the gap region between the low and middle corona. The significant overlap between the EUV and white-light features signifies that a leading shock wave may not have formed, or yet de-coupled from the CME, this early in the event. We conclude that the white-light front observed in K-Cor and LASCO C2 is the CME's pile-up compression region.

\section*{Acknowledgments}
This work is supported by the Bulgarian National Science Fund, VIHREN program, under contract KP-06-DV-8/18.12.2019 (MOSAIICS project). SDO/AIA data is courtesy of NASA/SDO and the AIA, EVE, and HMI science teams. K-Cor data is courtesy of the Mauna Loa Solar Observatory, operated by the High Altitude Observatory, as part of the National Center for Atmospheric Research (NCAR). NCAR is supported by the National Science Foundation.The SOHO/LASCO data used here are produced by a consortium of the Naval Research Laboratory (USA), Max-Planck-Institut fuer Aeronomie (Germany), Laboratoire d'Astronomie (France), and the University of Birmingham (UK). SOHO is a project of international cooperation between ESA and NASA. This research used version 4.0.5 of the SunPy open source software package \citep{Sunpy:2020}.

\begin{thebibliography}{65}
\providecommand{\natexlab}[1]{#1}
\providecommand{\url}[1]{\texttt{#1}}
\providecommand{\urlprefix}{URL }
\providecommand{\eprint}[2][]{\url{#2}}

\bibitem[{{Attrill} et~al.(2007){Attrill}, {Harra}, {van Driel-Gesztelyi}, and
  {D{\'e}moulin}}]{Attrill:2007}
{Attrill}, G.~D.~R., L.~K. {Harra}, L.~{van Driel-Gesztelyi}, and
  P.~{D{\'e}moulin}, 2007.
\newblock {Coronal ``Wave'': Magnetic Footprint of a Coronal Mass Ejection?}
\newblock \emph{The Astrophysical Journal Letters}, \textbf{656}, L101--L104.
\newblock 10.1086/512854.

\bibitem[{{Balmaceda} et~al.(2018){Balmaceda}, {Vourlidas}, {Stenborg}, and
  {Dal Lago}}]{Balmaceda:2018}
{Balmaceda}, L.~A., A.~{Vourlidas}, G.~{Stenborg}, and A.~{Dal Lago}, 2018.
\newblock {How Reliable Are the Properties of Coronal Mass Ejections Measured
  from a Single iewpoint?}
\newblock \emph{Astrophys. J.}, \textbf{863}(1), 57.
\newblock 10.3847/1538-4357/aacff8.

\bibitem[{Barnes et~al.(2020)Barnes, Cheung, Bobra, Boerner, Chintzoglou
  et~al.}]{Barnes:2020}
Barnes, W.~T., M.~C.~M. Cheung, M.~G. Bobra, P.~F. Boerner, G.~Chintzoglou,
  et~al., 2020.
\newblock aiapy: A Python Package for Analyzing Solar EUV Image Data from AIA.
\newblock \emph{Journal of Open Source Software}, \textbf{5}(55), 2801.
\newblock 10.21105/joss.02801,
  \urlprefix\url{https://doi.org/10.21105/joss.02801}.

\bibitem[{{Biesecker} et~al.(2002){Biesecker}, {Myers}, {Thompson}, {Hammer},
  and {Vourlidas}}]{Biesecker:2002}
{Biesecker}, D.~A., D.~C. {Myers}, B.~J. {Thompson}, D.~M. {Hammer}, and
  A.~{Vourlidas}, 2002.
\newblock {Solar Phenomena Associated with ``EIT Waves''}.
\newblock \emph{Astrophys. J.}, \textbf{569}(2), 1009--1015.
\newblock 10.1086/339402.

\bibitem[{{Brueckner} et~al.(1995){Brueckner}, {Howard}, {Koomen}, {Korendyke},
  {Michels} et~al.}]{Brueckner:1995}
{Brueckner}, G.~E., R.~A. {Howard}, M.~J. {Koomen}, C.~M. {Korendyke}, D.~J.
  {Michels}, et~al., 1995.
\newblock {The Large Angle Spectroscopic Coronagraph (LASCO)}.
\newblock \emph{Sol. Phys.}, \textbf{162}(1-2), 357--402.
\newblock 10.1007/BF00733434.

\bibitem[{Byrne et~al.(2013)Byrne, Long, Gallagher, Bloomfield, Maloney,
  McAteer, Morgan, and Habbal}]{byrne:2013}
Byrne, J., D.~Long, P.~Gallagher, D.~Bloomfield, S.~Maloney, R.~McAteer,
  H.~Morgan, and S.~R. Habbal, 2013.
\newblock Improved methods for determining the kinematics of coronal mass
  ejections and coronal waves.
\newblock \emph{Astronomy \& Astrophysics}, \textbf{557}, A96.

\bibitem[{{Cane} et~al.(1981){Cane}, {Erickson}, {Hanisch}, and
  {Turner}}]{Cane:1981}
{Cane}, H.~V., W.~C. {Erickson}, R.~J. {Hanisch}, and P.~J. {Turner}, 1981.
\newblock {Observations of rich clusters of galaxies at metre wavelengths.}
\newblock \emph{Monthly Notices of the Royal Astronomical Society},
  \textbf{196}, 409--415.
\newblock 10.1093/mnras/196.3.409.

\bibitem[{{Chen}(2006)}]{Chen:2006}
{Chen}, P.~F., 2006.
\newblock {The Relation between EIT Waves and Solar Flares}.
\newblock \emph{The Astrophysical Journal, Letters}, \textbf{641}(2),
  L153--L156.
\newblock 10.1086/503868.

\bibitem[{{Chen}(2009)}]{Chen:2009}
{Chen}, P.~F., 2009.
\newblock {The Relation Between EIT Waves and Coronal Mass Ejections}.
\newblock \emph{The Astrophysical Journal, Letters}, \textbf{698}(2),
  L112--L115.
\newblock 10.1088/0004-637X/698/2/L112, \eprint{0905.3272}.

\bibitem[{{Chen} et~al.(2002){Chen}, {Wu}, {Shibata}, and {Fang}}]{Chen:2002}
{Chen}, P.~F., S.~T. {Wu}, K.~{Shibata}, and C.~{Fang}, 2002.
\newblock {Evidence of EIT and Moreton Waves in Numerical Simulations}.
\newblock \emph{The Astrophysical Journal Letters}, \textbf{572}, L99--L102.
\newblock 10.1086/341486.

\bibitem[{{Cheng} et~al.(2020){Cheng}, {Zhang}, {Kliem}, {T{\"o}r{\"o}k},
  {Xing}, {Zhou}, {Inhester}, and {Ding}}]{Cheng:2020}
{Cheng}, X., J.~{Zhang}, B.~{Kliem}, T.~{T{\"o}r{\"o}k}, C.~{Xing}, Z.~J.
  {Zhou}, B.~{Inhester}, and M.~D. {Ding}, 2020.
\newblock {Initiation and Early Kinematic Evolution of Solar Eruptions}.
\newblock \emph{Astrophys. J.}, \textbf{894}(2), 85.
\newblock 10.3847/1538-4357/ab886a, \eprint{2004.03790}.

\bibitem[{{Dai} et~al.(2010){Dai}, {Auch{\`e}re}, {Vial}, {Tang}, and
  {Zong}}]{Dai:34}
{Dai}, Y., F.~{Auch{\`e}re}, J.~C. {Vial}, Y.~H. {Tang}, and W.~G. {Zong},
  2010.
\newblock {Large-scale Extreme-Ultraviolet Disturbances Associated with a Limb
  Coronal Mass Ejection}.
\newblock \emph{Astrophys. J.}, \textbf{708}(2), 913--919.
\newblock 10.1088/0004-637X/708/2/913.

\bibitem[{{Delaboudini{\`e}re} et~al.(1995){Delaboudini{\`e}re}, {Artzner},
  {Brunaud}, {Gabriel}, {Hochedez} et~al.}]{Delaboudini:1995}
{Delaboudini{\`e}re}, J.~P., G.~E. {Artzner}, J.~{Brunaud}, A.~H. {Gabriel},
  J.~F. {Hochedez}, et~al., 1995.
\newblock {EIT: Extreme-Ultraviolet Imaging Telescope for the SOHO Mission}.
\newblock \emph{Sol. Phys.}, \textbf{162}(1-2), 291--312.
\newblock 10.1007/BF00733432.

\bibitem[{{Fisher} and {Welsch}(2008)}]{Fisher:2008}
{Fisher}, G.~H., and B.~T. {Welsch}, 2008.
\newblock {FLCT: A Fast, Efficient Method for Performing Local Correlation
  Tracking}.
\newblock In R.~{Howe}, R.~W. {Komm}, K.~S. {Balasubramaniam}, and G.~J.~D.
  {Petrie}, eds., Subsurface and Atmospheric Influences on Solar Activity, vol.
  383 of \emph{Astronomical Society of the Pacific Conference Series}, 373.

\bibitem[{{Gallagher} and {Long}(2010)}]{Gallagher:2010}
{Gallagher}, P.~T., and D.~M. {Long}, 2010.
\newblock {Large-scale Bright Fronts in the Solar Corona: A Review of ''EIT
  waves''}.
\newblock \emph{Space Sci. Rev.}, 135--+.
\newblock 10.1007/s11214-010-9710-7.

\bibitem[{{Gopalswamy} et~al.(2000){Gopalswamy}, {Hanaoka}, and
  {Hudson}}]{Gopalswamy:2000}
{Gopalswamy}, N., Y.~{Hanaoka}, and H.~S. {Hudson}, 2000.
\newblock {Structure and Dynamics of the Corona Surrounding an Eruptive
  Prominence}.
\newblock \emph{Advances in Space Research}, \textbf{25}(9), 1851--1854.
\newblock 10.1016/S0273-1177(99)00597-9.

\bibitem[{Horn and Schunck(1981)}]{horn:1981}
Horn, B.~K., and B.~G. Schunck, 1981.
\newblock Determining optical flow.
\newblock \emph{Artificial Intelligence}, \textbf{17}(1), 185--203.
\newblock Https://doi.org/10.1016/0004-3702(81)90024-2,
  \urlprefix\url{https://www.sciencedirect.com/science/article/pii/0004370281900242}.

\bibitem[{{Hurlburt} et~al.(2012){Hurlburt}, {Cheung}, {Schrijver}, {Chang},
  {Freeland} et~al.}]{Hurlburt:2012}
{Hurlburt}, N., M.~{Cheung}, C.~{Schrijver}, L.~{Chang}, S.~{Freeland}, et~al.,
  2012.
\newblock {Heliophysics Event Knowledgebase for the Solar Dynamics Observatory
  (SDO) and Beyond}.
\newblock \emph{Sol. Phys.}, \textbf{275}(1-2), 67--78.
\newblock 10.1007/s11207-010-9624-2, \eprint{1008.1291}.

\bibitem[{{Hutton} and {Morgan}(2017)}]{HuttonMorgan:2017}
{Hutton}, J., and H.~{Morgan}, 2017.
\newblock {Automated detection of coronal mass ejections in three-dimensions
  using multi-viewpoint observations}.
\newblock \emph{Astron. Astrophys.p}, \textbf{599}, A68.
\newblock 10.1051/0004-6361/201629516, \eprint{1612.04560}.

\bibitem[{{Joshi} and {Srivastava}(2011)}]{JoshiSrivastava:2011}
{Joshi}, A.~D., and N.~{Srivastava}, 2011.
\newblock {Acceleration of Coronal Mass Ejections from Three-dimensional
  Reconstruction of STEREO Images}.
\newblock \emph{Astrophys. J.}, \textbf{739}(1), 8.
\newblock 10.1088/0004-637X/739/1/8, \eprint{1107.1769}.

\bibitem[{{Kim} et~al.(2019){Kim}, {Park}, {Lee}, {Moon}, {Bae}
  et~al.}]{Kim:2019}
{Kim}, T., E.~{Park}, H.~{Lee}, Y.-J. {Moon}, S.-H. {Bae}, et~al., 2019.
\newblock {Solar farside magnetograms from deep learning analysis of
  STEREO/EUVI data}.
\newblock \emph{Nature Astronomy}, \textbf{3}, 397--400.
\newblock 10.1038/s41550-019-0711-5.

\bibitem[{{Klassen} et~al.(2000){Klassen}, {Aurass}, {Mann}, and
  {Thompson}}]{Klassen:2000}
{Klassen}, A., H.~{Aurass}, G.~{Mann}, and B.~J. {Thompson}, 2000.
\newblock {Catalogue of the 1997 SOHO-EIT coronal transient waves and
  associated type II radio burst spectra}.
\newblock \emph{Astronomy \& Astrophysics, Supplement}, \textbf{141}, 357--369.
\newblock 10.1051/aas:2000125.

\bibitem[{{Kozarev} et~al.(2017){Kozarev}, {Davey}, {Kendrick}, {Hammer}, and
  {Keith}}]{Kozarev:2017}
{Kozarev}, K.~A., A.~{Davey}, A.~{Kendrick}, M.~{Hammer}, and C.~{Keith}, 2017.
\newblock The Coronal Analysis of SHocks and Waves (CASHeW) framework.
\newblock \emph{J. Space Weather Space Clim.}, \textbf{7}, A32.
\newblock 10.1051/swsc/2017028,
  \urlprefix\url{https://doi.org/10.1051/swsc/2017028}.

\bibitem[{{Kozarev} et~al.(2019){Kozarev}, {Dayeh}, and
  {Farahat}}]{Kozarev:2019}
{Kozarev}, K.~A., M.~A. {Dayeh}, and A.~{Farahat}, 2019.
\newblock {Early-stage Solar Energetic Particle Acceleration by Coronal Mass
  Ejection-driven Shocks with Realistic Seed Spectra. I. Low Corona}.
\newblock \emph{The Astrophysical Journal}, \textbf{871}(1), 65.
\newblock 10.3847/1538-4357/aaf1ce.

\bibitem[{{Kozarev} et~al.(2011){Kozarev}, {Korreck}, {Lobzin}, {Weber}, and
  {Schwadron}}]{Kozarev:2011}
{Kozarev}, K.~A., K.~E. {Korreck}, V.~V. {Lobzin}, M.~A. {Weber}, and N.~A.
  {Schwadron}, 2011.
\newblock {Off-limb Solar Coronal Wavefronts from SDO/AIA Extreme-ultraviolet
  Observations --- Implications for Particle Production}.
\newblock \emph{The Astrophysical Journal Letters}, \textbf{733}, L25.
\newblock 10.1088/2041-8205/733/2/L25.

\bibitem[{{Kozarev} et~al.(2015){Kozarev}, {Raymond}, {Lobzin}, and
  {Hammer}}]{Kozarev:2015}
{Kozarev}, K.~A., J.~C. {Raymond}, V.~V. {Lobzin}, and M.~{Hammer}, 2015.
\newblock {Properties of a Coronal Shock Wave as A Driver of Early SEP
  Acceleration}.
\newblock \emph{The Astrophysical Journal}, \textbf{799}(2), 167.
\newblock 10.1088/0004-637X/799/2/167.

\bibitem[{{Kozarev} and {Schwadron}(2016)}]{Kozarev:2016}
{Kozarev}, K.~A., and N.~A. {Schwadron}, 2016.
\newblock {A Data-driven Analytic Model for Proton Acceleration by Large-scale
  Solar Coronal Shocks}.
\newblock \emph{The Astrophysical Journal}, \textbf{831}, 120.
\newblock 10.3847/0004-637X/831/2/120.

\bibitem[{{Lemen} et~al.(2012){Lemen}, {Title}, {Akin}, {Boerner}, {Chou}
  et~al.}]{Lemen:2012}
{Lemen}, J.~R., A.~M. {Title}, D.~J. {Akin}, P.~F. {Boerner}, C.~{Chou},
  et~al., 2012.
\newblock {The Atmospheric Imaging Assembly (AIA) on the Solar Dynamics
  Observatory (SDO)}.
\newblock \emph{Solar Physics}, \textbf{275}, 17--40.
\newblock 10.1007/s11207-011-9776-8.

\bibitem[{{Li} and {Zhu}(2013)}]{Li:2013}
{Li}, R., and J.~{Zhu}, 2013.
\newblock {Solar flare forecasting based on sequential sunspot data}.
\newblock \emph{Research in Astronomy and Astrophysics}, \textbf{13}(9),
  1118-1126.
\newblock 10.1088/1674-4527/13/9/010.

\bibitem[{{Long} et~al.(2017){Long}, {Bloomfield}, {Chen}, {Downs}, {Gallagher}
  et~al.}]{Long:2017}
{Long}, D.~M., D.~S. {Bloomfield}, P.~F. {Chen}, C.~{Downs}, P.~T. {Gallagher},
  et~al., 2017.
\newblock {Understanding the Physical Nature of Coronal ``EIT Waves''}.
\newblock \emph{Sol. Phys.}, \textbf{292}(1), 7.
\newblock 10.1007/s11207-016-1030-y, \eprint{1611.05505}.

\bibitem[{{Long} et~al.(2014){Long}, {Bloomfield}, {Gallagher}, and
  {P{\'e}rez-Su{\'a}rez}}]{Long:2014}
{Long}, D.~M., D.~S. {Bloomfield}, P.~T. {Gallagher}, and
  D.~{P{\'e}rez-Su{\'a}rez}, 2014.
\newblock {CorPITA: An Automated Algorithm for the Identification and Analysis
  of Coronal ''EIT Waves''}.
\newblock \emph{Solar Physics}, \textbf{289}, 3279--3295.
\newblock 10.1007/s11207-014-0527-5.

\bibitem[{L{\"o}ptien et~al.(2016)L{\"o}ptien, Birch, Duvall, Gizon, and
  Schou}]{loptien_b_flct:2016}
L{\"o}ptien, B., A.~C. Birch, T.~L. Duvall, L.~Gizon, and J.~Schou, 2016.
\newblock Data compression for local correlation tracking of solar granulation.
\newblock \emph{Astronomy \& Astrophysics}, \textbf{587}, A9.

\bibitem[{Lucas and Kanade(1981)}]{lucas:1981}
Lucas, B.~D., and T.~Kanade, 1981.
\newblock An iterative image registration technique with an application to
  stereo vision.
\newblock In IJCAI'81: 7th international joint conference on Artificial
  intelligence, vol.~2, 674--679.

\bibitem[{{Majumdar} et~al.(2021){Majumdar}, {Patel}, {Pant}, and
  {Banerjee}}]{Majumdar:2021}
{Majumdar}, S., R.~{Patel}, V.~{Pant}, and D.~{Banerjee}, 2021.
\newblock {An Insight into the Coupling of CME Kinematics in Inner and Outer
  Corona and the Imprint of Source Regions}.
\newblock \emph{Astrophys. J.}, \textbf{919}(2), 115.
\newblock 10.3847/1538-4357/ac1592, \eprint{2107.08198}.

\bibitem[{McKenzie(2013)}]{mckenzie_flct:2013}
McKenzie, D., 2013.
\newblock Turbulent dynamics in solar flare sheet structures measured with
  local correlation tracking.
\newblock \emph{The Astrophysical Journal}, \textbf{766}(1), 39.

\bibitem[{{Mierla} et~al.(2008){Mierla}, {Davila}, {Thompson}, {Inhester},
  {Srivastava}, {Kramar}, {St. Cyr}, {Stenborg}, and {Howard}}]{Mierla:2008}
{Mierla}, M., J.~{Davila}, W.~{Thompson}, B.~{Inhester}, N.~{Srivastava},
  M.~{Kramar}, O.~C. {St. Cyr}, G.~{Stenborg}, and R.~A. {Howard}, 2008.
\newblock {A Quick Method for Estimating the Propagation Direction of Coronal
  Mass Ejections Using STEREO-COR1 Images}.
\newblock \emph{Sol. Phys.}, \textbf{252}(2), 385--396.
\newblock 10.1007/s11207-008-9267-8.

\bibitem[{{Moreton}(1960)}]{Moreton:1960}
{Moreton}, G.~E., 1960.
\newblock {H{$\alpha$} Observations of Flare-Initiated Disturbances with
  Velocities \~{}1000 km/sec.}
\newblock \emph{Astron. J.}, \textbf{65}, 494.
\newblock 10.1086/108346.

\bibitem[{{Moreton} and {Ramsey}(1960)}]{MoretonRamsey:1960}
{Moreton}, G.~E., and H.~E. {Ramsey}, 1960.
\newblock {Recent Observations of Dynamical Phenomena Associated with Solar
  Flares}.
\newblock \emph{Pub. Astron. Soc. Pacific}, \textbf{72}(428), 357.
\newblock 10.1086/127549.

\bibitem[{{Morgan} et~al.(2006){Morgan}, {Habbal}, and {Woo}}]{Morgan:2006}
{Morgan}, H., S.~R. {Habbal}, and R.~{Woo}, 2006.
\newblock {The Depiction of Coronal Structure in White-Light Images}.
\newblock \emph{Sol. Phys.}, \textbf{236}(2), 263--272.
\newblock 10.1007/s11207-006-0113-6, \eprint{astro-ph/0602174}.

\bibitem[{{Patel} et~al.(2021){Patel}, {Pant}, {Iyer}, {Banerjee}, {Mierla},
  and {West}}]{Patel:2021}
{Patel}, R., V.~{Pant}, P.~{Iyer}, D.~{Banerjee}, M.~{Mierla}, and M.~J.
  {West}, 2021.
\newblock {Automated Detection of Accelerating Solar Eruptions Using Parabolic
  Hough Transform}.
\newblock \emph{Sol. Phys.}, \textbf{296}(2), 31.
\newblock 10.1007/s11207-021-01770-z, \eprint{2010.14786}.

\bibitem[{{Patsourakos} et~al.(2009){Patsourakos}, {Vourlidas}, {Wang},
  {Stenborg}, and {Thernisien}}]{Patsourakos:2009a}
{Patsourakos}, S., A.~{Vourlidas}, Y.~M. {Wang}, G.~{Stenborg}, and
  A.~{Thernisien}, 2009.
\newblock {What Is the Nature of EUV Waves? First STEREO 3D Observations and
  Comparison with Theoretical Models}.
\newblock \emph{Solar Physics}, \textbf{259}, 49--71.
\newblock 10.1007/s11207-009-9386-x.

\bibitem[{{Pesnell} et~al.(2012){Pesnell}, {Thompson}, and
  {Chamberlin}}]{Pesnell:2012}
{Pesnell}, W.~D., B.~J. {Thompson}, and P.~C. {Chamberlin}, 2012.
\newblock {The Solar Dynamics Observatory (SDO)}.
\newblock \emph{Sol. Phys.}, \textbf{275}(1-2), 3--15.
\newblock 10.1007/s11207-011-9841-3.

\bibitem[{{Reiner} et~al.(1999){Reiner}, {Kaiser}, {Fainberg}, and
  {Stone}}]{ReinerKaiser:1999}
{Reiner}, M.~J., M.~L. {Kaiser}, J.~{Fainberg}, and R.~G. {Stone}, 1999.
\newblock {Remote radio tracking of CMEs in the solar corona and interplanetary
  medium}.
\newblock In S.~R. {Habbal}, R.~{Esser}, J.~V. {Hollweg}, and P.~A. {Isenberg},
  eds., Solar Wind Nine, vol. 471 of \emph{American Institute of Physics
  Conference Series}, 653--656.
\newblock 10.1063/1.58709.

\bibitem[{Sheeley~Jr et~al.(2014)Sheeley~Jr, Warren, Lee, Chung, Katz, and
  Namkung}]{Sheeley:2014}
Sheeley~Jr, N., H.~Warren, J.~Lee, S.~Chung, J.~Katz, and M.~Namkung, 2014.
\newblock Using running difference images to track proper motions of XUV
  coronal intensity on the Sun.
\newblock \emph{The Astrophysical Journal}, \textbf{797}(2), 131.

\bibitem[{{St. Cyr} et~al.(2017){St. Cyr}, {Posner}, and
  {Burkepile}}]{StCyr:2017}
{St. Cyr}, O.~C., A.~{Posner}, and J.~T. {Burkepile}, 2017.
\newblock {Solar energetic particle warnings from a coronagraph}.
\newblock \emph{Space Weather}, \textbf{15}(1), 240--257.
\newblock 10.1002/2016SW001545.

\bibitem[{{Stepanyuk} et~al.(2022){Stepanyuk}, {Kozarev}, and
  {Nedal}}]{Stepanyuk:2022}
{Stepanyuk}, O., K.~{Kozarev}, and M.~{Nedal}, 2022.
\newblock {Multi-scale image preprocessing and feature tracking for remote CME
  characterization}.
\newblock \emph{Journal of Space Weather and Space Climate}, \textbf{12}, 20.
\newblock 10.1051/swsc/2022020, \eprint{2205.15088}.

\bibitem[{Su et~al.(2013)Su, Veronig, Holman, Dennis, Wang, Temmer, and
  Gan}]{su_flct:2013}
Su, Y., A.~M. Veronig, G.~D. Holman, B.~R. Dennis, T.~Wang, M.~Temmer, and
  W.~Gan, 2013.
\newblock Imaging coronal magnetic-field reconnection in a solar flare.
\newblock \emph{Nature Physics}, \textbf{9}(8), 489--493.

\bibitem[{{Szenicer} et~al.(2019){Szenicer}, {Fouhey}, {Munoz-Jaramillo},
  {Wright}, {Thomas}, {Galvez}, {Jin}, and {Cheung}}]{Szenicer:2019}
{Szenicer}, A., D.~F. {Fouhey}, A.~{Munoz-Jaramillo}, P.~J. {Wright},
  R.~{Thomas}, R.~{Galvez}, M.~{Jin}, and M.~C.~M. {Cheung}, 2019.
\newblock {A deep learning virtual instrument for monitoring extreme UV solar
  spectral irradiance}.
\newblock \emph{Science Advances}, \textbf{5}(10), eaaw6548.
\newblock 10.1126/sciadv.aaw6548.

\bibitem[{{Telloni} et~al.(2022){Telloni}, {Zank}, {Stangalini}, {Downs},
  {Liang} et~al.}]{Telloni:2022}
{Telloni}, D., G.~P. {Zank}, M.~{Stangalini}, C.~{Downs}, H.~{Liang}, et~al.,
  2022.
\newblock {Observation of a Magnetic Switchback in the Solar Corona}.
\newblock \emph{Astrophys. J. Lett.}, \textbf{936}(2), L25.
\newblock 10.3847/2041-8213/ac8104, \eprint{2206.03090}.

\bibitem[{{Temmer} et~al.(2008){Temmer}, {Veronig}, {Vr{\v{s}}nak},
  {Ryb{\'a}k}, {G{\"o}m{\"o}ry}, {Stoiser}, and
  {Mari{\v{c}}i{\'c}}}]{Temmer:2008}
{Temmer}, M., A.~M. {Veronig}, B.~{Vr{\v{s}}nak}, J.~{Ryb{\'a}k},
  P.~{G{\"o}m{\"o}ry}, S.~{Stoiser}, and D.~{Mari{\v{c}}i{\'c}}, 2008.
\newblock {Acceleration in Fast Halo CMEs and Synchronized Flare HXR Bursts}.
\newblock \emph{The Astrophysical Journal, Letters}, \textbf{673}(1), L95.
\newblock 10.1086/527414.

\bibitem[{{The SunPy Community} et~al.(2020){The SunPy Community}, Barnes,
  Bobra, Christe, Freij et~al.}]{Sunpy:2020}
{The SunPy Community}, W.~T. Barnes, M.~G. Bobra, S.~D. Christe, N.~Freij,
  et~al., 2020.
\newblock The SunPy Project: Open Source Development and Status of the Version
  1.0 Core Package.
\newblock \emph{The Astrophysical Journal}, \textbf{890}, 68--.
\newblock 10.3847/1538-4357/ab4f7a,
  \urlprefix\url{https://iopscience.iop.org/article/10.3847/1538-4357/ab4f7a}.

\bibitem[{{Thernisien} et~al.(2009){Thernisien}, {Vourlidas}, and
  {Howard}}]{Thernisien:2009}
{Thernisien}, A., A.~{Vourlidas}, and R.~A. {Howard}, 2009.
\newblock {Forward Modeling of Coronal Mass Ejections Using STEREO/SECCHI
  Data}.
\newblock \emph{Sol. Phys.}, \textbf{256}(1-2), 111--130.
\newblock 10.1007/s11207-009-9346-5.

\bibitem[{{Thompson} et~al.(1998){Thompson}, {Plunkett}, {Gurman}, {Newmark},
  {St.~Cyr}, and {Michels}}]{Thompson:1998}
{Thompson}, B.~J., S.~P. {Plunkett}, J.~B. {Gurman}, J.~S. {Newmark}, O.~C.
  {St.~Cyr}, and D.~J. {Michels}, 1998.
\newblock {SOHO/EIT observations of an Earth-directed coronal mass ejection on
  May 12, 1997}.
\newblock \emph{Geophysical Review Letters}, \textbf{25}, 2465--2468.
\newblock 10.1029/98GL50429.

\bibitem[{{Tomczyk} et~al.(2022){Tomczyk}, {Burkepile}, {De Wijn}, {Gibson},
  {Gilbert} et~al.}]{Tomczyk:2022}
{Tomczyk}, S., J.~{Burkepile}, A.~{De Wijn}, S.~{Gibson}, H.~{Gilbert}, et~al.,
  2022.
\newblock {The Coronal Solar Magnetism Observatory}.
\newblock In The Third Triennial Earth-Sun Summit (TESS, vol.~54,
  2022n7i121p01.

\bibitem[{{Tripathi} and {Raouafi}(2007)}]{Tripathi:2007}
{Tripathi}, D., and N.~E. {Raouafi}, 2007.
\newblock {On the relationship between coronal waves associated with a CME on 5
  March 2000}.
\newblock \emph{Astron. Astrophys.p}, \textbf{473}(3), 951--957.
\newblock 10.1051/0004-6361:20077255.

\bibitem[{{Uchida}(1968)}]{Uchida:1968}
{Uchida}, Y., 1968.
\newblock {Propagation of Hydromagnetic Disturbances in the Solar Corona and
  Moreton's Wave Phenomenon}.
\newblock \emph{Sol. Phys.}, \textbf{4}(1), 30--44.
\newblock 10.1007/BF00146996.

\bibitem[{{Uchida}(1974)}]{Uchida:1974}
{Uchida}, Y., 1974.
\newblock {Behaviour of the Flare-Produced Coronal MHD Wavefront and the
  Occurrence of Type II Radio Bursts}.
\newblock \emph{Sol. Phys.}, \textbf{39}(2), 431--449.
\newblock 10.1007/BF00162436.

\bibitem[{{Veronig} et~al.(2010){Veronig}, {Muhr}, {Kienreich}, {Temmer}, and
  {Vr{\v{s}}nak}}]{Veronig:2010}
{Veronig}, A.~M., N.~{Muhr}, I.~W. {Kienreich}, M.~{Temmer}, and
  B.~{Vr{\v{s}}nak}, 2010.
\newblock {First Observations of a Dome-shaped Large-scale Coronal
  Extreme-ultraviolet Wave}.
\newblock \emph{The Astrophysical Journal, Letters}, \textbf{716}(1), L57--L62.
\newblock 10.1088/2041-8205/716/1/L57, \eprint{1005.2060}.

\bibitem[{Vourlidas et~al.(2013)Vourlidas, Lynch, Howard, and
  Li}]{Vourlidas:2013}
Vourlidas, A., B.~J. Lynch, R.~A. Howard, and Y.~Li, 2013.
\newblock How many CMEs have flux ropes? Deciphering the signatures of shocks,
  flux ropes, and prominences in coronagraph observations of CMEs.
\newblock \emph{Solar Physics}, \textbf{284}(1), 179--201.

\bibitem[{{Vrsnak} et~al.(1995){Vrsnak}, {Ruzdjak}, {Zlobec}, and
  {Aurass}}]{Vrsnak:1995}
{Vrsnak}, B., V.~{Ruzdjak}, P.~{Zlobec}, and H.~{Aurass}, 1995.
\newblock {Ignition of MHD Shocks Associated with Solar Flares}.
\newblock \emph{Sol. Phys.}, \textbf{158}(2), 331--351.
\newblock 10.1007/BF00795667.

\bibitem[{{Wang} et~al.(2009){Wang}, {Vahala}, and {Vahala}}]{Wang:2009}
{Wang}, T., G.~{Vahala}, and L.~{Vahala}, 2009.
\newblock {Scalar Magnetic Field Distribution Function Approach to MHD
  Turbulence}.
\newblock In APS Division of Plasma Physics Meeting Abstracts, vol.~51 of
  \emph{APS Meeting Abstracts}, BP8.057.

\bibitem[{{Webb} and {Howard}(2012)}]{Webb:2012}
{Webb}, D.~F., and T.~A. {Howard}, 2012.
\newblock {Coronal Mass Ejections: Observations}.
\newblock \emph{Living Reviews in Solar Physics}, \textbf{9}, 3.

\bibitem[{{Welsch} et~al.(2004){Welsch}, {Fisher}, {Abbett}, and
  {Regnier}}]{welsch_flct:2004}
{Welsch}, B.~T., G.~H. {Fisher}, W.~P. {Abbett}, and S.~{Regnier}, 2004.
\newblock {ILCT: Recovering Photospheric Velocities from Magnetograms by
  Combining the Induction Equation with Local Correlation Tracking}.
\newblock \emph{The Astrophysical Journal}, \textbf{610}(2), 1148--1156.
\newblock 10.1086/421767.

\bibitem[{{Wild} and {McCready}(1950)}]{WildMcCready:1950}
{Wild}, J.~P., and L.~L. {McCready}, 1950.
\newblock {Observations of the Spectrum of High-Intensity Solar Radiation at
  Metre Wavelengths. I. The Apparatus and Spectral Types of Solar Burst
  Observed}.
\newblock \emph{Australian Journal of Scientific Research A Physical Sciences},
  \textbf{3}, 387.
\newblock 10.1071/CH9500387.

\bibitem[{{Zhang} and {Dere}(2006)}]{Zhang:2006}
{Zhang}, J., and K.~P. {Dere}, 2006.
\newblock {A Statistical Study of Main and Residual Accelerations of Coronal
  Mass Ejections}.
\newblock \emph{Astrophys. J.}, \textbf{649}(2), 1100--1109.
\newblock 10.1086/506903.

\end{thebibliography}

\end{document}